\documentclass[12pt]{article}

\newtheorem{Theorem}{Theorem}[section]

\newtheorem{Proposition}[Theorem]{Proposition}

\makeatletter
\usepackage{latexsym}
\usepackage[T1]{fontenc}
\usepackage{amsmath}
\usepackage{amssymb}

\def \AO {{\cal A}({\cal O})}
\def \AO' {{\cal A}({\cal O}')}

\def \] {\supseteq}

\def \be {\begin{equation}}
\def \ee {\end{equation}}

\def \ra {\rightarrow}

\def \eqq {\equiv}

\def \b {{\beta}}

\def \eps {{\varepsilon}}

\def \l {{\lambda}}

\def \ph {{\varphi}}

\def \A {{\cal A}}
\def \B {{\cal B}}
\def \C {{\cal C}}
\def \D {{\cal D}}
\def \F {{\cal F}}
\def \G {{\cal G}}
\def \H {\mbox{${\cal H}$}}

\def \K {{\cal K}}

\def \O {{\cal O}}

\def \T {{\cal T}}
\def \U {{\cal U}}

\def \Z {{\cal Z}}

\def \Psio {{\Psi_0}}

\def \x {{\bf x}}

\def \Rbf {{\bf R}}

\def \AO {{\cal A}({\cal O})}
\def \AO' {{\cal A}({\cal O}')}

\def \] {\supseteq}

\newfam\Ssfam
\font\eleSs=cmss10 at12pt \font\sevenSs= cmss10 at 8pt \font\sixSs= cmss10 at 6pt

\textfont\Ssfam=\eleSs\scriptfont\Ssfam=\sevenSs%
\scriptscriptfont\Ssfam=\sixSs \def\Ss{\fam\Ssfam\eleSs}

\def\doppio#1{{\rm I}\kern-.1667em{\rm #1}}

\def\Q{\text{Q}\kern-.52em
    \text{\vrule height1.5ex width.5pt depth0pt}\kern.45em}

\def\Z{{\mathchoice {\hbox{$\Ss\textstyle Z\kern-0.4em Z$}}
{\hbox{$\Ss\textstyle Z\kern-0.4em Z$}} {\hbox{$\Ss\scriptstyle Z\kern-0.25em
Z$}} {\hbox{$\Ss\scriptscriptstyle Z\kern-0.2em Z$}}}}

\def\C{{\mathchoice{\hbox{$\rm\textstyle\text{\kern.35em\vrule
   height1.5ex width.5pt depth0pt\kern-.35em C}$}}
{\hbox{$\rm\textstyle\text{\kern.35em\vrule
   height1.5ex width.5pt depth0pt\kern-.35em C}$}}
{\hbox{$\rm\scriptstyle\text{\kern.35em\vrule
   height1.5ex width.3pt depth0pt\kern-.35em C}$}}
{\hbox{$\rm\scriptscriptstyle\text{\kern.35em\vrule
   height1.5ex width.2pt depth0pt\kern-.35em C}$}}}}

\def \be{\begin{equation} \displaystyle}
\def \ee{\end{equation}}

\def \A*{\mbox{$A^{*} $}}
\def \B*{\mbox{$B^{*} $}}
\def \C*{\mbox{$C^{*} $}}

\def \bea{\begin{eqnarray}}
\def \eea{\end{eqnarray}}

\def \b{\beta}

\def \l{\lambda}

\@addtoreset{equation}{section}

\def \be {\begin{equation} \displaystyle}

\def \ee {\end{equation}}

\def \ra {\rightarrow}
\def\AO {\mbox{${\cal A}({\cal O})$}}
\def\AO'{\mbox{${\cal A}({\cal O}')$}}
\def\O {\mbox{${\cal O}$}}

\def\A{\mbox{${\cal A}$}}

\def \ra{\rightarrow}
\def \ph {{\varphi}}
\def \eps {{\varepsilon}}

\def \O {{\cal O}}

\def \A {{\cal A}}
\def \AO {\A(\O)}
\def \AOl'{\A(\O_{loc}')}

\def \B {{\cal B}}
\def \F {{\cal F}}
\def \D {{\cal D}}

\def \H {{\cal H}}

\def \x {{\bf x}}

\begin{document}
%\begin{titlepage}
  \title{Some Rigorous  Results on Symmetry Breakings  in Gauge QFT{\footnote{Invited contribution to the XXXIII International Workshop on High Eenergy Physics "Hard Problems of Hadron Physics: Non-Perturbative QCD and Related Quests" Nov. 8-12, 2021, Logunov Institute of High Energy Physics, Russia}}}

\sloppy

\author{F. Strocchi \\Dipartimento di Fisica, Universit\`a di Pisa,
 Pisa, Italy  }

%\end{titlepage}

\fussy

\date{}

\maketitle

\begin{abstract}
The extraordinary success of the Standard Model asks  for a more rigorous control beyond the perturbative approach, which is affected by mathematical problems (interaction picture  and canonical quantization, non-covergence of the perturbative series, triviality results for $\phi^4$ model and related models, etc.).  
   We shall briefly discuss some non-perturbative results concerning the crucial role and realization of symmetry breakings in the Standard Model.  

\end{abstract}

%\end{titlepage}

\section{BEH mechanism}
A crucial structural ingredient of the standard model is the BEH (briefly Higgs) mechanism, which has been discussed and used at the perturbative level with an expansion based on a mean field ansatz. Since mean field expansions are known to give incorrect results  for the critical temperature and the energy spectrum in spin models, one would like to control the mechanism with a rigorous non-perturbative approach. This will also free the conclusions from  the weak points of the perturbative expansion, which is known to lead to a non convergent series and relies on the interaction picture and  canonical quantization,  both  mathematically excluded for non-trivial interactions.\footnote{For a discussion of the mathematical problems of the perturbative expansion see F. Strocchi, \textit{ An Introduction to Non-Perturbative Foundations of Quantum Field Theory}, Oxford University Press 2013, 2016, hereafter referred to as F. Strocchi [2016]. }
   
As it is well known, but sometime overlooked in the textbook discussions of the Higgs mechanism, in order to avoid the exclusion of a symmetry  breaking order parameter by Elitzur theorem, the first crucial  step it to introduce a gauge fixing;  then, the discussion of the way the mechanism avoids the occurrence of massless Goldstone bosons becomes gauge fixing dependent.  
 Furthermore, in view of a non-pertubative approach,  a mean field term, which breaks the global gauge group, should not appear in the gauge fixing, because   it  would   require an \textit{a priori} non-perturbative  control of its selfconsistency. One then considers gauge fixings invariant under the global gauge group.\footnote{The motivations for such strategic choices, which exclude the pathological unitary gauge as well as the so-called $\xi$ gauges, are discussed in F. Strocchi [2016].} 

    From a rigorous point if view, one faces the problem of proving that the spontaneous  breaking of the global gauge group does not imply the existence of massless Goldstone bosons. This shall be dealt with in the BRST gauge and, in the abelian case, in the Coulomb gauge.
\vspace{2mm}

\noindent i) {\bf{\textit{Absence of Goldstone particles in the Higgs mechanism}}}

We choose to discuss the problem in the BRST gauge.  The advantage is that the corresponding  field algebra $\F$ is local, so that one may control the generation of the infinitesimal transformations, under  the global gauge group, by conserved local currrents, $J^a_\mu$, as in the proof of the Goldstone Theorem, thanks to the non-renormalization theorem for the commutators of local conserved currents with local  fields:
\be{ < \delta^a F > = \lim_{R \ra \infty} < [ Q^a_R, F ] >,  \,\,\,\,\forall F \in \F,}\ee
where  $Q^a_R$ denotes the suitably regularized charge localized in the sphere of radius $R$
\be{ Q^a_R = J^a_0(f_R \alpha) = \int d^4 x \,  J^a_0(x) f_R (\x) \, \alpha(x_0),}\ee
  $ f_R(\x) = f(|\x|/R), \, f(x) = 1$, for $|x| \leq R$, $f(x) = 0$, for $|x| > R(1 + \eps)$,\,  supp\,$\alpha \,\subseteq [- \eps, \eps]$, $\tilde{\alpha}(0) = \int d x_0\, \alpha(x_0) =1$,  $f, \alpha\, \in C^\infty$.  

 \begin{Theorem} {\textit{(Higgs mechanism)}}
In the BRST quantization of a Yang-Mills theory, the spontaneous breaking of a one-parameter subgroup of the global gauge group $G$  by the vacuum expectation of $F \in \F$, $ < \delta^a F > \,\neq 0,$ implies the existence of  a $\delta(k^2)$ singularity in the Fourier transform of $< F \,J^a_\mu(x)  >$, ({\bf \em  massless  
Goldstone modes} in the  $a$-channel), where $J^a_\mu(x) $ is the conserved current which generates the infinitesimal transformations of the local fields under such  a one-parameter  subgroup;
  however, such  modes cannot describe physical particles.
\end{Theorem}
 The basic ingredient for the proof is the validity of the Local Gauss Law, on the physical states, which in the BRST gauge reads
 \be{ J_\mu^a = \partial^\nu F_{\mu \nu}^a + \{\,Q_B, \,(D_\mu \bar{c})^a\,\},}\ee
where  $Q_B$ denotes the BRST charge and $\bar{c}$ one of the ghost fields.\footnote{For details see S. Weinberg, \textit{The Quantum Theory of Fields, Vol. II,}, Cambridge University Press 1996, Sect. 15.7; F. Strocchi [2016], Chapter 7, Section 4.}    
The physical state vectors $\Psi$ are selected by the BRST subsidiary condition
\be{ Q_B \,\Psi = 0,\,\,\,\,\,\,\,\,\,\Rightarrow  \,\,\,\,\,< \Psi, (J^a_\mu - \partial^\nu F_{\mu \nu}^a)\,\Psi > \,= 0.}\ee 
One may prove that this excludes a contribution to such a massless mode by  a physical state,  as intermediate state  in the two-point function  $< F \,J^a_\mu(x)  >$.\footnote{G.De Palma and F. Strocchi, Ann. Phys. {\bf{336}}, 112 (2013); F. Strocchi, \textit{Symmetry Breaking}, Springer, third edition 2021, Appendix E}.

The unphysical nature of the massless modes in local renormalizable gauges has been argued  within a perturbative expansion.\footnote{See the very comprehensive review: G. Gurlanik, C.R. Hagen and T.W. Kibble, Broken symmetries and the Goldstone theorem, in {\em Advances in particle Physics, Vol.\,2}, R.L. Good and R.E. Marshak eds., Interscience 1968.} The above non perturbative result
 improves the perturbative  analysis, since it does not rely on a semiclassical mean field ansatz nor on the summability of the perturbative series; moreover the order parameter is not restricted to be a pointlike field. 

%Furthermore, the exclusion of physical Goldstone bosons associated to the breaking of the global gauge group $G$, provides a sharper argument with respect to the mere  non-applicability of the Goldstone theorem in physical gauge by the lack of local order parameters, which may invalidate the local generation of the symmetry.

\vspace{2mm}

\noindent  ii) {\bf{\textit{A theorem on the abelian Higgs mechanism}}}

In the abelian case one may prove a sharper result,\footnote{G. Morchio and F. Strocchi, Jour. Phys. A: Math. Phys.{ \bf{40}}, 3173 (2007); F. Strocchi [2016], Chapter 7, Section 6.2;  \textit{Symmetry Breaking in the Standard Model. A non-perturbative outlook}, Scuola Normale Superiore, 2019, Section 2.8.} which, in particular,  includes  a rigorous link between the disappearance of the Goldstone boson and the vector boson becoming massive (beyond the popular anthropomorphic picture in unphysical gauges).

This is obtained in the Coulomb gauge; the advantage  is that the corresponding field algebra $\F_C$ does not contain unphysical fields and all the states of its Hilbert space representation have a physical meaning. On the other hand, the validity of the Local Gauss Law implies that $\F_C$ is  non local; as a consequence, one looses the   control of the local generation of the infinitesimal  transformations
$\delta^{U(1)}$ of the fields under the global $U(1)$ gauge group $\b^\l, \l \in \Rbf$:
\be{ \delta^{U(1)} F \eqq  \frac{d \b^\l(F)}{ d \l}|_{\l = 0}, \,\,\,F \in \F_C.}\ee
Since the charged fields $F$, characterized by $\delta^{U(1)} F \neq 0$, are non-local,  the non-renormalization theorem for the commutators of conserved currents does not apply, and in fact, contrary what  is usually taken for granted, one may prove that          the commutatators of the space integral of the current charge density $j_0(\x, x_0)$ are not independent of the  time $x_0$ 
\be{ \lim_{R \ra \infty} [\,j_0(f_R, x_0), \,\ph^C(y)\,] = - e \int d m^2\, \rho(m^2)\,\cos(m(x_0 - y_0))\,\ph^C(y),   }\ee
 where  $\ph^C$ is the Coulomb charged field and $\rho(m^2)$ is the spectral measure which defines the two point function of the vector boson field $F_{\mu \nu}$
\be{< F_{\mu \nu}(x)\,F_{\l\,\sigma}(y) > = i d_{\mu \nu \l \sigma} \int d m^2 \rho(m^2) \,\Delta^+(x-y; m^2),}\ee
$$  d_{\mu \nu \l \sigma} = g_{\nu \sigma} \partial_\mu \partial_\l + g_{\mu \l} \partial_\nu \partial_\sigma - g_{\nu \l} \partial_\mu \partial_\sigma - g_{\mu \sigma} \partial_\nu \partial_\l.$$
 
 In order to find a relation between the current charge density $j_0(f_R, x_0)$ and the electric charge, at least in the unbroken case, an improved smearing is needed\footnote{M. Requardt, Commun. Math. Phys. {\bf 50}, 259 (1976); G. Morchio and F. Strocchi, Jour. Math. Phys. {\bf 44}, 5569 (2003).} which amounts to  introducing
$Q_{ \delta R} \eqq j_0(f_R \alpha_{\delta R}), \,\alpha_{\delta R} \eqq \alpha(x_0/\delta R)/\delta R,$ $0 < \delta < 1$ ($\alpha$ as in eq.\,(1.2)).
Then, the so obtained charge has  time independent commutators and annihilates the vacuum
\be{ \delta_c F \eqq  i \lim_{ \delta \ra 0, \,R \ra \infty} [\,Q_{\delta R}, \, F\,],\,\,\,\,\,\,\,\lim_{R \ra \infty} Q_{\delta  R} \Psio = 0. }\ee

The next question is the relation between the derivation $\delta_c$,  induced by the current charge density  and the derivation $\delta^{U(1)}$. 
Such a relation turns out to play a crucial role for the following general theorem on the Higgs phenomenon.

\begin{Theorem} ({\bf Higgs phenomenon})

\noindent {\bf A.} The current and the $U(1)$  derivations  coincide, $\delta_c = \delta^{U(1)}$,  if and only if the two point spectral function of 
the vector field $F_{\mu \nu}$ contains a $\delta(m^2)$, namely if the corresponding   {\bf  vector boson is massless};  in this case, the global {\bf $U(1)$ is unbroken} and the matrix elements of its generator $Q$ are given by 
\be{ <\Psi,\, Q\,\Phi > = \lim_{\delta \ra 0, R \ra \infty} < \Psi, \,Q_{\delta R}\,\Phi >,}\ee 
(for all the Coulomb states $\Psi,\,\Phi$). Thus, thanks to the improved smearing, one recovers the expected  {\bf \em relation between the charge density and the $U(1)$ charge},  although in an alerted form.

\vspace{1mm}
\noindent {\bf B.} The global {\bf $U(1)$ gauge group is broken}, i.e. $ < \delta^{U(1)}\,F  > \neq 0$, $F \in \F_C$,   only if  $ \delta^{U(1)} \neq \delta_c $ and in this case,
%\vspace{1mm} 

\noindent i) the  {\bf   vector boson is  massive}; 
%\vspace{1mm}

\noindent ii)  the Goldstone spectrum, defined by the Fourier transform of the two point function $< j_0(x)\,F > $, is governed by the spectral function of the vector field, and therefore  cannot contain any $\delta(k^2)$ (i.e. there are {\bf no  associated Goldstone bosons});

%\vspace{1mm}
\noindent  iii) the Gauss charge, defined by the suitably smeared flux of $F_{0 \,i}$ at space infinity, vanishes on the Coulomb states ({\bf screening of the Gauss charge}):
\be{ \lim_{\delta \ra 0, \,R \ra \infty}  < \Psi, \,Q_{\delta R}\,\Phi > = 0.}\ee
 \end{Theorem}

%%%%%%%%%%%%%%%%%%%%%%%%%%%%%%%%%%%%%%%%%%%%%%%%
%%%%%%%%%%%%%%%%%%%%%%%%%%%%%%%%%%%%%%%%%%%%%%

\section{Gauge group topology solves the $U(1)$ problem and yields the $\theta$ vacuum structure}

A corner stone for the control of QCD structure is  the discovery of the role of  topology, yielding  the $\theta $ vacuum structure and a  solution of the $U(1)$ problem.  The standard treatment is based  on the assumption that  the  euclidean functional integral is governed by the instanton solutions (semiclassical approximation), which are classified by their topological winding number $n$. The functional integral is thus  evaluated by first integrating over the class of euclidean (continuous) configurations with given {\em winding number} $n$ and then by summing over $n$, with a weighting factor  $e^{ i \,\theta \,n}$, where $\theta$ is a {\em free} parameter, the so-called $\theta$ {\em angle}.

Such a procedure is not free of mathematical problems, since, already in the free field case, the set of continuous euclidean configurations has zero functional measure. Hence, in contrast with the quantum mechanical case,  the WKB (semiclasssical) approximation is problematic in QFT. 

Furthermore, such an approach does not clearly settle the debated question of whether  the axial $U(1)_A$ transformations may be still defined for the observable fields,  so that $U(1)_A$ is \textit{spontaneously} broken in QCD.

A rigorous solution of such problems may be obtained following a suggestion by Roman Jackiw.\footnote{G. Morchio and  F. Strocchi, Ann. Phys. {\bf{324}}, 2236 (2009); F. Strocchi, \textit{Symmetry Breaking}, third edition,  Springer 2021, and references therein.}
The idea is to directly exploit the non-trivial topology of the gauge group, rather than its reflexes on the classification of the instanton solutions; in this way one does not make any reference to the  problematic semiclassical instanton approximation. 
\vspace{1mm}

\noindent ii) {\bf{\textit{Solution of the $U(1)$ problem }}}

This is conveniently obtained in the temporal gauge, which is local and positive;  the only delicate (but mathematically crucial) point is that, as a consequence of the required invariance of the vacuum under the Local Gauss Law operator,  the represented  field algebra $\F$  is generated by the gauge invariant fields and by the formal exponentials of the gauge dependent fields (with algebraic relations corresponding to those of their formal exponentials). 

The first step is the definition of time independent  $U(1)_A$  transformations $\b^\l, \l \in \Rbf$ of  $\F$ and, in particular, of its observable subalgebra $\F_{obs}$
\be{ \b^\l(F) = \lim_{R \ra \infty} V^5_R(\l)\,F\,V_R^5(-\l),\,\,\,\,\,\,\,\forall F \in \F,}\ee
where the one-parameter  unitary operators $V_R^5(\l)$ are (formally) the exponentials $e^{ i  \l\,J_0^5(f_R \alpha)}$, with $J^5_\mu$ the conserved (gauge dependent) current
\be{ J^5_\mu = j^5_\mu  - (16  \pi^2)^{-1} \eps_{\mu \nu \rho \sigma}  \mbox{Tr}\, [ F^{\nu \rho}\, A^\sigma - (2/3) A^\nu A^\rho A^\sigma ] \eqq j_\mu^5 + K_\mu^5,}\ee
( $j^5_\mu$ is the gauge invariant anomalous current); by locality, the limit is reached for finite $R$. 

The gauge dependence of the   unitary operators $V^5_R(\l)$ does not invalidate the above definition, since they are  merely  instrumental for the definition of the chiral transformations $\b^\l$ on the observable fields, a result which  is clearly independent of the gauge fixing and of the corresponding (gauge dependent) field algebra in which $\F_{obs}$ is embedded.   It looks short sighted\footnote{G. t' Hooft, How instantons solve the $U(1)$ problem, Physics Reports, {\bf 142}, 357 (1986).} to blame on the fact that $J^5_\mu$ or the (better behaved)  exponentials  $V^5_R(\l)$ are gauge dependent  not  observable operators; such a point of view would in fact deny the very existence of the non-abelian gauge symmetries of the standard model, being generated by gauge dependent currents. 

Given the {\em existence of the $U(1)_A$ transformations of the observable fields},  a gauge independent fact, no matter how its actual existence is proved, the real issue is the  mechanism for evading the Goldstone theorem, for which the non-abelianess of the gauge group should play a decisive role.

The next step is the interplay between  axial transformations and gauge transformations. 
To this purpose we analyse the properties of the   local gauge group $\G$, left unbroken by the gauge fixing, with elements  
$\alpha_\U$ parametrized by time independent 
$C^\infty$ unitary functions $\U(\x)$, taking value in the global group $\mathbb{G}$ and   differing from the identity only on a compact set,  $\K_\U \subset \Rbf^3$. Thanks to their  space localization the $\U$ obviously extend to the one-point compactification of $\Rbf^3$, $\dot{{\Rbf}}^3$, which is isomorphic to the three sphere $S^3$, and define {\em continuous} mappings of $S^3$ onto the global gauge group $\mathbb{G}$:
$ \U(\x): \,\,\, \dot{{\Rbf}}^3\, \sim S^3\,\,\ra \,\,\mathbb{G}.$
Such mappings $\U$ fall into disjoint homotopy classes labeled by the (topological invariant) {\bf winding number} $n(\U)$  
\be{ n(\U) = (24 \pi^2)^{-1} \int d^3 x \,\eps^{i j k} \,\mbox{Tr}\,[ \U_i(\x)\,\U_j(\x)\,\U_k(\x) ] \eqq \int d^3 x \,n_\U(\x), }\ee     $\U_i(\x) \eqq \U(\x)^{-1} \partial_i\,\U(\x)$;  $\U_n$  shall denote a  function with winding number $n$. 

The one-parameter groups  of unitary gauge functions
$$\U(\l g) = e^{i \l \,g(\x)},\,\,\,\,\l \in \Rbf,\,\,\,\, g(\x) = g_a(\x)\,T^a, \,\,\,\,g_a \in \D(\Rbf^3),$$ ($T^a$ the representative matrices of the generators of $\mathbb{G}$) continuously connected to the identity,  define  a subgroup  $\G_0 \subset \G$, which   is generated  by  the unitary operators $V(\U(\l g)) \in \F$,  formally the exponentials of the Gauss operator $G^a \eqq ({\bf D}\cdot {\bf E})^a - j^a_0, \,\,j_\mu^a = i \bar{\psi} \,\gamma_\mu t^a \psi,$
$V(\U(\l g)) \sim e^{i \l\,G(g)}, \,\,G(g) \eqq \sum_a G^a(g_a),\,\,g_a \in \D(\Rbf^3), $
 since formally $ \delta^{g_a} F = i \,[\, G^a(g^a),\,F\,]$, (the operators $V(\U(\l g)) $ need not to be represented by weakly continuous unitary operators).   $\G_0$ is called the \textit{ Gauss subgroup} of $\G$ and its elements have zero winding number. In the following, for simplicity, we shall often adopt the short-hand notation $\U(g)$, or $\U_g$.

 In the Hilbert space  $\H$ defined by the correlation functions of a vacuum state  $\omega_0$, the   physical state vectors $\Psi$  are selected by the subsidiary condition
\be{ V(\U(\l g)) \Psi = \Psi, \,\,\,\,\,\,\forall \,\,\U(\l g) \in \G_0, \,\,\,\,\,\, \Psi \in \H' \subset \H.}\ee
By exploiting the localization property of the gauge functions and the locality of $\F$, one shows that the state $\omega_0$ is invariant under the full group $\G$, namely $\omega_0(\alpha_\U(F)) = \omega_0(F),\, \forall F \in \F$,  so that $\G$ is implemented by unitary operators $V(\U)$ in $\H$ and
   \be{ V(\U_n)\, V^5_R(\l)\,  V(\U)^{-1} = e^{ i \l\,2 n}\, V^5_R(\l).}\ee

\begin{Proposition}   The spontaneous breaking of the  $U(1)_A$ symmetry $\b^\l$ in QCD, by $< \delta^5 A > \neq 0$, with $ \delta^5 A$ the infinitesimal $U(1)_A$ transformation of $A$ and $A$ an observable (hermitian) field, evades the Goldstone theorem because $\delta^5 A$ cannot be  related to the two point function of $A$ and a (local conserved) current, as required for the proof of  the Goldstone .
\end{Proposition}
The point is that, as a consequence of the non-trivial topology of $\G$ and of the above equations, one has  
$$ < V^5_R(\l)\, A > = < \alpha_{\U_n} (V^5_R(\l)\, A) > =  e^{i 2 n \l}\, < V^5_R(\l)\, A >.$$
This proves that $< V^5_R(\l)\, A >$ is a singular function of $\l$ and its derivative with respect to $\l$ does not exists; then, even if the axial $U(1)_A$ transformations are given by the action of the local unitary operators $ V^5_R(\l)$, their  infinitesimal form  cannot be given by commutators  with a local  conserved current, here   $J^5_\mu$, a crucial assumption for the proof of the Goldstone theorem. 

\vspace{1,5mm}
\noindent  ii) {\bf{\textit{Gauge topological group and $\theta$ vacuum structure}}}

The  topology of the gauge group $\G$  is  described by the quotient of $\G$ by its normal subgroup $\G_0$.  $\T \eqq \G/\G_0$, is an abelian group with elements $\T_n$ which are classified by the (topological) winding number $n$, and  commute with the gauge transformations.

\begin{Theorem} In the Hilbert space  representation of the temporal gauge field algebra $\F$, by a (Gauss invariant) vacuum state $\omega_0$, the topological group $\T$ is represented by gauge invariant operators $T_n$, which belong to the center of the algebra of observables $\A$ and reduce to unitary operators  on the space  $\H'$ of physical states, with spectrum $e^{i\,2n \,\theta}, \, \theta \in [0, \pi)$. 
    Thus, the representations of the observable algebra in the physical space are labelled by the angle $\theta$.

Under $U(1)_A$  transformations $\b^\l$, one has 
\be{\b^\l(T_n) = e^{i\, 2n\,\l}\, T_n,}\ee
so that in each  representation of $\A$ with trivial center, $U(1)_A$ is {\bf {always}}  broken.

If the Gauss invariant vacuum state $\omega$ defines an irreducible representation of the field algebra $\F$, then it selects a definite value of $\theta$
\be{ T_n\, \Psio = e^{i\, 2n\,\theta}\, \Psio.}\ee
\end{Theorem}

Thus, the $\theta$ angle arises in an intrinsic way as a label of the spectrum of the center of the algebra of observables, uniquely selected  in each irreducible (vacuum) representation of the field algebra, rather than as a free parameter in the semiclassical instanton approximation of the euclidean functional integral.

 \end{document}